\documentclass{amsart}  
\usepackage{amssymb}  
\usepackage{latexsym}  

\newcommand{\Z}{{\mathbb Z}}  
\newcommand{\R}{{\mathbb R}}  
\newcommand{\C}{{\mathbb C}}  
\newcommand{\N}{{\mathbb N}}

\newtheorem{theorem}{Theorem}     
     
\newtheorem{lemma}{Lemma}[section]     
\newtheorem{prop}[lemma]{Proposition}     
\newtheorem{coro}[lemma]{Corollary}     
     
\newcommand{\tr}{{\mathrm{tr}}}  
  
\newcounter{smalllist}

\begin{document}  
\title[Quasi-Sturmian potentials]{Uniform spectral properties of one-dimensional quasicrystals, IV.\ Quasi-Sturmian potentials}  
\author[D.~Damanik, D.~Lenz]{David Damanik$\,^{1}$ and Daniel Lenz$\,^{2,3}$*}\thanks{* Research of D.~L.  was supported in part by THE ISRAEL SCIENCE
FOUNDATION (grant no. 447/99) and   the Edmund Landau Center for Research in Mathematical Analysis and Related Areas, sponsored by the Minerva Foundation (Germany).}
\maketitle
\vspace{0.3cm}
\noindent
$^1$ Department of Mathematics, University of California, Irvine, CA 92697, USA\\[2mm]      
$^2$ Institute of Mathematics, The Hebrew University, Jerusalem 91904, Israel\\
$^3$ Fachbereich Mathematik, Johann Wolfgang Goethe-Universit\"at,      
60054 Frankfurt, Germany\\[2mm]
E-mail: \mbox{damanik@math.uci.edu, dlenz@math.uni-frankfurt.de}\\[3mm]
1991 AMS Subject Classification: 81Q10, 47B80\\      
Key words: Schr\"odinger operators, low-complexity potentials, singular continuous spectrum, zero-measure Cantor spectrum, Hausdorff
dimensional spectral properties

\begin{abstract}
We consider discrete one-dimensional Schr\"odinger operators with quasi-Sturmian potentials. We present
a new approach to the trace map dynamical system which is independent of the initial conditions and
establish a characterization of the spectrum in terms of bounded trace map orbits. Using this, it is
shown that the operators have purely singular continuous spectrum and their spectrum is a Cantor set of
Lebesgue measure zero. We also exhibit a subclass having purely $\alpha$-continuous spectrum. All these
results hold uniformly on the hull generated by a given potential.

\end{abstract}

\section{Introduction}

This article studies spectral properties of discrete one-dimensional Schr\"odinger operators

\begin{equation}\label{oper}
(H \phi) (n) = \phi(n+1) + \phi(n-1) + V(n) \phi(n)
\end{equation}
in $\ell^2(\Z)$ with potential $V : \Z \rightarrow \R$ of low combinatorial complexity. Consider the
case where $V$ takes finitely many values. A particular example is given by the case where $V$ is
periodic. It is well known that periodicity of $V$ implies that the spectrum of $H$ is purely
absolutely continuous. Among the aperiodic cases, several classes of potentials have been studied.
These include, for example, Sturmian potentials \cite{bist,d1,dkl,dl1,dl2}, potentials generated by
circle maps \cite{dp,hks}, potentials generated by substitutions
\cite{bbg,bg,d5,d6,hks}, and random (Bernoulli-type) potentials \cite{ckm}.

There are two important observations that can be made. In all the known results, the spectral type is
pure, that is, we do not know of any example within the class of operators with potential taking
finitely many values that has mixed spectral type. On the other hand, any spectral type can occur:
Periodic potentials always lead to purely absolutely continuous spectrum, Sturmian potentials always
lead to purely singular continuous spectrum \cite{dkl}, and almost all random potentials lead to pure
point spectrum \cite{ckm}. Potentials generated by circle maps or substitutions seem to always lead to
purely singular continuous spectrum; at least no counterexample is known yet.

A point of view that was raised in \cite{d2} and will be more comprehensively discussed in \cite{ad}
is the following: 

\begin{equation}\label{tend}
\mbox{The more complex the potential, the more singular the spectral type.}
\end{equation}

Here, complexity is to be understood in a combinatorial sense. By restricting $V$ to finite intervals,
we can speak of ``finite subwords'' of
$V$. The (combinatorial) complexity function $p : \N \rightarrow \N$ then assigns to each $n \in \N$ the number $p(n)$ of finite subwords of $V$
that have length $n$. As it is well known in the combinatorics on words community, periodicity of $V$
is characterized by $p$ being bounded and Sturmian $V$ are (essentially) characterized by the fact that
$p$ has lowest possible growth among the unbounded complexity functions. On the other hand, random
potentials have almost surely maximal complexity in the sense that if $V$ takes values in $\mathcal{A}
\subseteq \R$ and is generated randomly, then almost surely, $p(n) = |\mathcal{A}|^n$ for every $n \in
\N$. Thus, reformulating \eqref{tend} as

\begin{equation}\label{tend2}
\mbox{The faster $p(n)$ grows, the more singular the spectral type.}
\end{equation}
we see that \eqref{tend2} holds true for the extreme cases, that is, maximal complexity growth,
minimal complexity growth (no growth), and minimal complexity growth among the unbounded cases. Between
Sturmian and random potentials, only few results are known. In particular it is not clear where the
transition from singular continuous spectrum to pure point spectrum occurs in terms of complexity.
Results in \cite{ds} show that one can construct potentials with very high complexity that lead to
purely singular continuous spectrum. 

A natural next step is to approximate the transition from above and below, that is, to extend
\cite{dkl} and \cite{ckm}. Our purpose here is to do the former and we investigate the next natural
complexity class: Quasi-Sturmian potentials. We will show that they, too, always lead to purely
singular continuous spectrum. 

It is an intriguing fact that the final result is of a very deterministing nature (every
quasi-Sturmian operator has purely singular continuous spectrum), yet the proofs make crucial use of
probabilistic methods by employing Kotani's theory for ergodic families of operators. It would be
interesting to see whether one can do away with this and give a deterministic proof in the same
generality. 

Apart from Kotani theory, another crucial input is a detailed study of the orbits of a certain dynamical
system, the quasi-Sturmian trace map, which is the heart of this paper. These results rely only on
certain hierarchical structures and are to a large extent model-independent. They may thus be of of
independent interest. Moreover, when specialized to the Sturmian case, they provide a different proof
of a central result in \cite{bist}.

Using our trace map results, we can show that the spectrum of Schr\"odinger operators with
quasi-Sturmian potentials is always purely singular continuous; and in some cases it has even stronger
continuity properties. 

As we will see later, quasi-Sturmian potentials, while always finitely valued, can take arbitrarily many
values, whereas Sturmian potentials take only two values. In this sense, the class of Sturmian
potentials is much richer, in particular in terms of their local properties which can in fact be
arbitrary.

This paper is a continuation of \cite{dkl,dl1,dl2} (cf.\ \cite{len} as well) and it further exploits
the general theme of the series: Application of canonical partitions of potentials to spectral theory
of the induced operators. Therefore, the three cornerstones of the theory are Kotani theory, trace map
bounds, and partitions. Whenever one can establish these three pieces for a given class of models, one
can obtain a good understanding of the behavior of generalized eigenfunctions of the operators and
hence their spectral properties.

The article is organized as follows. In Section~\ref{models} we describe in detail the models we
discuss and the results we obtain. The heart of the paper is Section~\ref{central} where we investigate
the quasi-Sturmian trace map and characterize the energies from the spectrum in terms of trace map
behavior. Sections~\ref{ac} and~\ref{pp} establish the singular continuous spectral type and
Section~\ref{alpha} refines this from the point of view of Hausdorff dimensional properties. In the
appendix we discuss some combinatorial properties of quasi-Sturmian sequences which we need in our
proofs.

\section{Models and Results}\label{models}

Let $\mathcal{A}$ be a finite set, called alphabet, and let $\mathcal{A}^*$ denote the set of finite
words built from elements of $\mathcal{A}$. Given a word $w \in \mathcal{A}^*$, we define the length
$|w|$ to be the number of symbols it is built from, that is, $|b_1
\ldots b_n| = n$. The elements of $\mathcal{A}^\N,\mathcal{A}^\Z$ are called one-sided (resp.,
two-sided) infinite sequences over
$\mathcal{A}$. Given a word or infinite sequence $u$ over $\mathcal{A}$, any finite subword of it is
called a factor. We denote by $F(u)$ the set of all factors of $u$.

Given a word or infinite sequence $u$, we define for $n \ge 1$,
$$
p_u(n) = \# \{ \mbox{ factors of $u$ having length $n$ } \}.
$$
The function $p_u : \N \rightarrow \N$ is called factor complexity (or just complexity) of $u$. If $u$
is a one-sided infinite sequence, then the following fundamental result holds \cite{hm}:

\begin{prop}[Hedlund-Morse]
The following are equivalent:

\begin{itemize}
\item[(i)] $u$ is eventually periodic {\rm (}i.e., there exist $k,n_0$ with $u(n+k) = u(n)$ for $n \ge n_0${\rm )}.
\item[(ii)] $p_u$ is bounded.
\item[(iii)] There exists $n_0$ with $p(n_0) \le n_0$.
\end{itemize}
\end{prop}

This shows that the complexity function displays a dichotomy. It is either bounded (if $u$ is eventually periodic) or it grows at least
linearly (if $u$ is aperiodic) with universal lower bound $p_u(n) \ge n+1$ for every $n$.

In the following we will only consider sequences $u$ that are recurrent, that is, every one of its
subwords occurs infinitely often. We will not always make this assumption explicit but we will give a
remark below which explains why these are the ``interesting cases.''

Recurrent sequences $u$ with minimal complexity $p_u(n) = n+1$ for every $n$ exist and they are called
Sturmian. More generally, a recurrent $u$ is called quasi-Sturmian if there are $k,n_0$ with $p_u(n) =
n+k$ for $n \ge n_0$.

From a combinatorial point of view, among the aperiodic sequences, Sturmian and quasi-Sturmian
sequences are the closest to eventually periodic sequences. This has led mathematical physicists to
consider their associated hulls (to be defined below) as standard models of one-dimensional
quasicrystals (see \cite{sbgc} for the discovery of quasicrystals and \cite{d2} for a survey of their
spectral theory in one dimension).

Given a one-sided infinite sequence $u$ over $\mathcal{A}$, we define its associated hull (also called
induced subshift) by
$$
\Omega_u = \{ \omega \in \mathcal{A}^\Z : \mbox{ every factor of $\omega$ is a factor of $u$ } \}.
$$
If we define the shift transformation $T : \mathcal{A}^Z \rightarrow \mathcal{A}^\Z$ by $(Tx)(n) =
x(n+1)$ and endow $\mathcal{A}$ with discrete topology and $\mathcal{A}^\Z$ with product topology, then
$(\Omega_u, T)$ is a topological dynamical system. If $u$ is uniformly recurrent (i.e., every factor
occurs infinitely often and with bounded gaps), $(\Omega_u, T)$ is minimal. If the frequencies of
factors exist uniformly (see \cite{q} for definitions and details), $(\Omega_u, T)$ is uniquely
ergodic, that is, there exists a unique $T$-invariant measure $\mu$ which is necessarily ergodic. A
subshift that is both minimal and uniquely ergodic is called strictly ergodic. It is known, and in fact
can be shown using the methods we present in the appendix, that a quasi-Sturmian sequence $u$ induces a
strictly ergodic subshift $\Omega_u$.

Given a subshift $(\Omega, T)$ and an injective function $f : \mathcal{A} \rightarrow \R$, we define,
for $\omega \in \Omega$, potentials $V_\omega : \Z \rightarrow \R$ by $V_\omega(n) = f(\omega(n))$. This
gives rise to a family $\{H_\omega\}_{\omega \in \Omega}$ of discrete one-dimensional Schr\"odinger
operators in $\ell^2(\Z)$ defined by
$$
(H_\omega \phi) (n) = \phi(n+1) + \phi(n-1) + V_\omega(n) \phi(n),
$$
where $\omega \in \Omega$ and $\phi \in \ell^2(\Z)$. If the subshift $\Omega$ is minimal, it follows
by a strong approximation argument that the spectrum of $H_\omega$ is deterministic, that is, there
exist a compact set $\Sigma \subseteq \R$ such that

\begin{equation}\label{spectrum}
\sigma(H_\omega) = \Sigma \mbox{ for every } \omega \in \Omega.
\end{equation}

Our first main result is the following.

\begin{theorem}\label{main}
Suppose $u$ is quasi-Sturmian and $f$ is one-to-one. Then for every $\omega \in \Omega_u$, the
operator $H_\omega$ has purely singular continuous spectrum and its spectrum is a Cantor set of
Lebesgue measure zero.
\end{theorem}

\noindent\textit{Remark.} We do not really need that $f$ is one-to-one. All results in this article
hold under the weaker assumption that $f$ is such that the resulting potentials $V_\omega$ are
aperiodic. In fact, most results hold for arbitrary $f$. The only place where we need aperiodicity of
the potentials is in Section~\ref{ac} where we apply Kotani theory. Of course, for periodic potentials,
the results in Sections~\ref{pp} and~\ref{alpha} are well known.

\medskip

This result has been known only in the Sturmian case and even there it was completed only very
recently after a 1989 paper by Bellissard et al.\ had already established zero-measure spectrum and
hence absence of absolutely continuous spectrum \cite{bist}. To establish uniform absence of eigenvalues
\cite{dkl}, it has proved useful to employ combinatorial notions and methods; see \cite{dl1} in
particular.

The present paper pushes the approach of \cite{dkl,dl1,dl2,len} further and extends their result and
that of \cite{bist} to all quasi-Sturmian cases. Not only is this a larger class, with still many
claims to model quasicrystalline structures in one dimension; as we will see below, this
class comprises models with arbitrary alphabet size, whereas Sturmian models are always based on a
two-letter alphabet (since $p(1) = 1+1 = 2$) which seems to be a physically ill-motivated restriction.

We will also be able to extend the other main result of \cite{dkl} to the quasi-Sturmian case. Namely,
the operators in question have always purely $\alpha$-continuous spectrum for some strictly positive
$\alpha$ which is uniform on the hull, provided that the underlying rotation number has bounded
density; see below for a precise statement of the result and Section~\ref{alpha} for the
proof. This implies in particular that all spectral measures associated with the operators $H_\omega$
give zero weight to sets with zero $\alpha$-dimensional Hausdorff measure and one gets quantum
dynamical implications by applying, for example, \cite{l}; see \cite{d1,dkl} for details.

Let us briefly comment on the case of non-recurrent quasi-Sturmian sequences. Using arguments similar
to the ones used in the appendix one can relate this class to the class of non-recurrent Sturmian
sequences. This class is in turn well understood \cite{ch} and contains only sequences which are
eventually periodic and hence are not natural candidates for quasicrystal models.

Next we want to employ a result of Cassaigne (originally due to Coven and Paul) to establish for
quasi-Sturmian sequences a weak analogue to the partitions of Sturmian sequences found in \cite{dl1}
(cf.\ \cite{len2} as well). Before we state Cassaigne's result, we recall the following. Let
$\mathcal{A},\mathcal{A}'$ be alphabets. A map $\mathcal{A} \rightarrow (\mathcal{A}')^*$ is called a
substitution (or morphism). It can be morphically extended to $\mathcal{A}^*$ (resp., $\mathcal{A}^\N$)
by $S(b_1 \ldots b_n) = S(b_1) \ldots S(b_n)$ (resp., $S(b_1 b_2 \ldots) = S(b_1) S(b_2) \ldots$). A
substitution on a two-letter alphabet (say, $\mathcal{A} = \{a,b\}$) is called aperiodic if $S(ab)
\not= S(ba)$. The following proposition, proved in \cite{c}, establishes a characterization of
quasi-Sturmian sequences in terms of substitutive images of Sturmian sequences.

\begin{prop}\label{cass}
A one-sided sequence $u \in \mathcal{A}^\N$ is quasi-Sturmian if and only if there exist a word $w \in
\mathcal{A}^*$, a Sturmian sequence $u_{{\rm St}} \in \{a,b\}^\N$, and an aperiodic substitution $S :
\{a,b\} \rightarrow \mathcal{A}^*$ such that
\begin{equation}\label{casseq}
u = w S(u_{{\rm St}}).
\end{equation}
\end{prop}

\noindent\textit{Remark.} While we base our presentation and discussion of Proposition~\ref{cass}
on~\cite{c}, we would like to point out that the result has been known for a long time. It was shown in
the papers \cite{c2} by Coven and \cite{p} by Paul.

\medskip

In order to appreciate the usefulness of the above proposition, let us recall that the local structure
of Sturmian sequences is very well understood \cite{b}: Given a Sturmian sequence $u_{{\rm St}}$, there
is a unique irrational number $\theta \in (0,1)$ such that

\begin{equation}\label{faceq}
F(u_{{\rm St}}) = F(c_\theta),
\end{equation}
where $c_\theta$ is given by

\begin{itemize}
\item $s_{-1} = a$, $s_0 = b$, $s_1 = s_0^{a_1 - 1}s_{-1}$, and 

\begin{equation}\label{snrec}
s_n = s_{n-1}^{a_n} s_{n-2}
\end{equation}
for $n \ge 2$, where the $a_n$ are the coefficients of the continued fraction expansion of $\theta$,
\item $c_\theta = \lim_{n \rightarrow \infty} s_n$ in the sense that $s_n$ is a prefix of $s_{n+1}$
for every $n \ge 1$ and $|s_n| \rightarrow \infty$.
\end{itemize}

For a given quasi-Sturmian sequence, it is therefore natural to associate with it the following set of
words,

\begin{equation}\label{sn'}
s_n' = S(s_n), \; n \ge -1.
\end{equation}

By virtue of \eqref{casseq} and \eqref{faceq}, it appears natural to carry over the partition approach
to Sturmian sequences developed in \cite{dl1} (cf.\ \cite{len2} as well), which decomposes a given
sequence into blocks of type $s_n$ and $s_{n-1}$ (for every $n$), to the quasi-Sturmian case. In fact,
we will use a weak analogue of the partition lemma from \cite{dl1} to study quasi-Sturmian potentials.

We saw above that we can associate an irrational $\theta$ with a given quasi-Sturmian sequence $u$.
Any such $\theta$ will be called a rotation number of $u$. We show in the appendix that there are in
general multiple choices for $\theta$, but the proof of Proposition~\ref{cass}, which will be sketched
in the appendix, suggests a particular choice for $\theta$, which will be called
\textit{canonical rotation number} and denoted by $\theta_c$. Consider the continued fraction
expansion of $\theta_c$,

$$
\theta_c = \cfrac{1}{a_1+ \cfrac{1}{a_2+ \cfrac{1}{a_3 + \cdots}}}
$$
with uniquely determined $a_n \in \N$. The associated rational approximants $\frac{p_n}{q_n}$ are
defined by 

\begin{alignat*}{3}
p_0 &= 0, &\quad	p_1 &= 1,   &\quad	p_n &= a_n p_{n-1} + p_{n-2},\\
q_0 &= 1, &		q_1 &= a_1, &		q_n &= a_n q_{n-1} + q_{n-2}.
\end{alignat*}
The number $\theta_c$ has bounded density if
$$
\limsup_{n \rightarrow \infty} \frac{1}{n} \sum_{i=1}^n a_i < \infty.
$$
The set of bounded density numbers is uncountable but has Lebesgue measure zero.

Now we are in position to state our second main result.

\begin{theorem}\label{acont}
Suppose $u$ is quasi-Sturmian with canonical rotation number $\theta_c$ having bounded density and $f$
is one-to-one. Then there exists $\alpha > 0$ such that for every $\omega \in \Omega$, the operator
$H_\omega$ has purely $\alpha$-continuous spectrum.
\end{theorem}

We finish this section with a discussion of certain symmetry  properties of quasi-Sturmian models.
These symmetry properties come from reflection symmetry of the underlying Sturmian dynamical systems.
The results below  will be used in Section \ref{alpha}. They extend the corresponding results of
\cite{dl2,len} to our setting. 

\medskip

For a word $w=w_1\ldots w_n$ over $\mathcal{A}=\{a,b\}=\{0,1\}$, we define the reflected word $w^R$ by $w^R = w_n
\ldots w_1$. Similarly, for a two-sided infinite word $\omega$ over $\mathcal{A}$, we define
$\omega^R$ by $\omega^R(n) = \omega(-n)$. It is well known that for arbitrary $\theta$, the
Sturmian dynamical system $\Omega(\theta)$ associated to $c_\theta$ is reflection invariant, that is,
it satisfies $\Omega(\theta) = \Omega(\theta)^R = \{ \omega^R : \omega \in \Omega(\theta)\}$.  

A proof can be given as  follows: By \cite{b}, for example, the $s_n$ satisfy
\begin{equation}\label{palin}
s_{2k +1}  = \pi_{2k +1} 01, \:\;  s_{2k} = \pi_{2k} 1 0, \: k \in \N,
\end{equation}
with suitable palindromes $\pi_n$ (i.e., $\pi_n=\pi_n^R$). The discussion above shows that
$\Omega(\theta)$ is determined by the set of $s_n$,
$n\geq 2$, in the sense that a double-sided-infinite sequence $\omega$ over $\{0,1\}$ belongs to 
$\Omega(\theta)$ if and only if every factor  of $\omega$ is a factor of a (suitable) $s_n$. By
\eqref{palin}, we see that $\Omega(\theta)$  is determined in this sense by the set of
$\pi_n$ as well. As  every $\pi_n$ is a palindrome, we infer  $\Omega(\theta) = \Omega(\theta)^R$. 

Now, let $\Omega(\theta, S)$ be the quasi-Sturmian dynamical system associated to $c_\theta$ and the
substitution $S$ according to Proposition \ref{cass}. Let $S^R$ be the substitution on $\{0,1\}$ with
$S^R (0) = S(0)^R$ and $S^R (1) = S(1)^R$. 

Mimicking the reasoning leading to $\Omega(\theta) = \Omega(\theta)^R$, we directly infer the
following proposition.

\begin{prop}\label{reflection}
$\Omega(\theta,S)^R = \Omega(\theta, S^R)$. 
\end{prop}

A direct calculation shows that $U_R H_\omega U_R^*= H_{\omega^R}$ holds for every
double-sided-infinite  $\omega$ over $\mathcal{A}$. Here, $U_R$ is the unitary reflection operator in
$\ell^2$ given by $(U_R u) (n) = u(-n)$.  Thus, the proposition immediately gives the following
corollary.

\begin{coro}\label{invarianz}
Denote by $\Sigma(\Omega))$ the deterministic spectrum of random operators associated to $\Omega$.
Then, $\Sigma(\Omega(\theta,S))=\Sigma(\Omega(\theta,S^R)$. 
\end{coro}

\section{The Quasi-Sturmian Trace Map}\label{central}

In this section we discuss the trace map associated with a quasi-Sturmian hull. We will show that the
spectrum coincides with the set of energies for which the corresponding trace map orbit remains bounded
and we give uniform bounds for these orbits. This result will be crucial to what follows and it allows
us to pursue a strategy similar to \cite{dkl,dl1,dl2,len}. 

Before we actually begin our discussion, let us emphasize the following: While our results on trace maps are similar to the corresponding results in the
Sturmian case treated in \cite{bist}, our proofs are completely different and in fact both more general and more conceptual. This is necessary as the
corresponding investigation of \cite{bist} cannot be carried over to quasi-Sturmian potentials (see below for details).

Recall that a quasi-Sturmian hull $\Omega = \Omega_u$ comes with a sequence of words $s_n'$, $n \ge
-1$, defined in \eqref{sn'}. If $s_n'$ has the form $s_n' = b_1 \ldots b_k$ with $b_i \in \mathcal{A}$
(leaving the dependence of $k$ and the $b_i$'s on $n$ implicit) and $E \in \C$, we define the transfer
matrix $M_E(n)$ by

\begin{equation}\label{mndef}
M_E(n) = \left( \begin{array}{cc} E - f(b_k) & -1 \\ 1 & 0 \end{array} \right) \times \cdots \times \left( \begin{array}{cc} E - f(b_1) & -1
\\ 1 & 0 \end{array} \right).
\end{equation}
It follows from \eqref{snrec} and \eqref{sn'} that for every $E \in \C$ and every $n \ge 2$,

\begin{equation}\label{mnrec}
M_E(n) = M_E(n-2) M_E(n-1)^{a_n}.
\end{equation}
In particular, the matrices $M_E(n)$ satisfy the same recursive relations as the transfer matrices in
the Sturmian case; compare \cite{bist}. Thus, their traces

\begin{equation}\label{xndef}
\tau_E(n) = \tr(M_E(n))
\end{equation}
satisfy the Sturmian recursive relations as well, however, with possibly different initial conditions.
This presents a potential difficulty. The Sturmian trace map preserves a quantity known as the
Fricke-Vogt invariant. In the Sturmian case, the initial conditions are always such that $E$ drops out,
that is, the invariant is $E$-independent. This may not be the case in our more general setting.
However, since the invariant is given by a polynomial in $E$, and the spectrum is compact, it is
uniformly bounded for energies from the spectrum. This fact will enable us to prove results similar to
the Sturmian case. Another important remark is in order: The proof of the classification of trace map
orbits in the Sturmian case \cite{bist} makes crucial use of a certain property of the initial
conditions and hence does not extend to the quasi-Sturmian case where this property may not hold. We
will therefore give a different proof of a similar classification result which works without any
assumptions on the initial conditions.

When studying the traces $\tau_E(n)$ defined in \eqref{xndef}, it is convenient to introduce the
following variables:

\begin{equation}\label{xyzdef}
( x_E(n) , y_E(n) , z_E(n) ) = ( \tfrac{1}{2} \tr(M_E(n-1)) , \tfrac{1}{2} \tr(M_E(n)) , \tfrac{1}{2}
\tr(M_E(n) M_E(n-1)) ).
\end{equation}

Using \eqref{mnrec} and the Cayley-Hamilton theorem, it is possible to find for every $n \in \N$, a
polynomial $F_n$ from $\R^3$ to itself such that for every $E,n$, we have

\begin{equation}
F_n(x_E(n), y_E(n), z_E(n)) = (x_E(n+1), y_E(n+1), z_E(n+1)).
\end{equation}
Namely, with the Chebyshev polynomials $U_m(x)$ (defined by $U_{-1} = 0$, $U_0 = 1$,
$U_{m+1}(x) = 2x U_m(x) - U_{m-1}(x)$), we have

\begin{equation}\label{tm}
F_n (x,y,z) = (y, z U_{a_{n+1}-1} (y) - x U_{a_{n+1}-2}(y), z U_{a_{n+1}} (y) - x U_{a_{n+1}-1}(y)).
\end{equation}

We want to study the orbits $(F_n \cdots F_1 (x,y,z))_{n \in \N}$ for arbitrary initial vectors
$(x,y,z) \in \R^3$. Our goal is to show that they are either bounded or super-exponentially diverging
in every coordinate. Moreover, we want to show that the spectrum consists of exactly those energies $E$
for which the corresponding orbit $(( x_E(n) , y_E(n) , z_E(n) ))_{n \in \N}$ is bounded. We will
employ methods from combinatorial group theory which were introduced by Roberts in \cite{r}; see
\cite{mks} for background and \cite{d3} for extensions of \cite{r} and further applications.

Let us first recall several known results. We refer the reader to \cite{r} (and references therein).
Each $F_n$ preserves the Fricke-Vogt invariant, defined by

\begin{equation}
I(x,y,z) = x^2 + y^2 + z^2 - 2xyz -1.
\end{equation}
The set $\mathcal{M}$ of all such polynomial mappings with complex coefficients,

\begin{equation}
\mathcal{M} = \{ F \in \C [x,y,z]^3 : I(F(x,y,z)) = I(x,y,z) \},
\end{equation}
is a group and can be written as a semidirect product of two groups,

\begin{equation}
\mathcal{M} = \Sigma \otimes_s \mathcal{G},
\end{equation}
where 

\begin{equation}
\Sigma = \{ \sigma_0 = Id, \sigma_1, \sigma_2, \sigma_3\} 
\end{equation}
is the normal subgroup. The involutions $\sigma_i$ are the pairwise sign changes, for example,
$\sigma_1(x,y,z) = (x,-y,-z)$. The group
$\mathcal{G}$ can be generated by the involution $p$ and the infinite-order mapping $u$, where

\begin{equation}
p(x,y,z) = (y,x,z), \; u(x,y,z) = (z,y,2yz - x).
\end{equation}
For $F_n$ from \eqref{tm}, we find

\begin{equation}\label{tmrep}
F_n = p u^{a_{n+1}}.
\end{equation}

As a group, $\mathcal{G}$ is isomorphic to the projective linear group ${\rm PGl}(2,\Z)$. The subgroup
$\mathcal{G}_{{\rm OP}} \sim {\rm PSl}(2,\Z)$ of orientation-preserving elements can be written as 

\begin{equation}
\mathcal{G}_{{\rm OP}} = \langle v \rangle \times \langle q \rangle,
\end{equation}
where

\begin{equation}
v(x,y,z) = (y,x,2xy - z), \; q(x,y,z) = (y,z,x).
\end{equation}
Using

\begin{equation}
u = vq, \; p u^k p = (v q^{-1})^k,
\end{equation}
we get the sequence $G_n$ of orientation-preserving elements, defined by

\begin{equation}\label{qvrep}
G_n = F_{2n} F_{2n-1} = (v q^{-1})^{a_{2n+1}} (v q)^{a_{2n}}.
\end{equation}
Let

\begin{equation}\label{undef}
U_n = G_n \cdots G_1,
\end{equation} 
so that

\begin{equation}
U_n (x_E(1), y_E(1), z_E(1)) = (x_E(2n+1), y_E(2n+1), z_E(2n+1)).
\end{equation}

Thus the $U_n$--orbit corresponds to trace map iterates with odd index. We get the trace map iterates
with even index in a similar way: Let

\begin{equation}\label{qvrep2}
H_n = F_{2n+1} F_{2n} = (v q^{-1})^{a_{2n+2}} (v q)^{a_{2n+1}}.
\end{equation}
With

\begin{equation}\label{vndef}
V_n = H_n \cdots H_1,
\end{equation} 
so that

\begin{equation}
V_n (x_E(2), y_E(2), z_E(2)) = (x_E(2n+2), y_E(2n+2), z_E(2n+2)).
\end{equation}

The following proposition is our central result on trace map orbits. It shows that unboundedness of
orbits is equivalent to entry in the \textit{escape set}

\begin{equation}
\mathcal{E} = \{ (x,y,z) \in \R^3 : |y| > 1, |z| > 1, |yz| > |x| \}.
\end{equation}
Moreover, once an orbit enters $\mathcal{E}$, it remains there and diverges super-exponentially in
every coordinate. This exhibits a dichotomy in trace map orbit behavior. A trace map orbit is either
bounded or superexponentially diverging, and one has in fact explicit bounds on the norms of iterates
when the orbit is bounded.

\begin{prop}\label{orbitchar}
For each energy $E$, the following are equivalent:

\begin{itemize}
\item[{\rm (i)}] $(U_n(x_E(1),y_E(1),z_E(1)))_{n \in \N}$ is unbounded. 
\item[{\rm (ii)}] There exists $N \in \N$ such that $U_N(x_E(1),y_E(1),z_E(1))) \in \mathcal{E}$.
\item[{\rm (iii)}] There exists $N \in \N$ such that $U_n(x_E(1),y_E(1),z_E(1))) \in \mathcal{E}$ for
each $n \ge N$.
\item[{\rm (iv)}] $(U_n(x_E(1),y_E(1),z_E(1))))_{n \in \N}$ is super-exponentially diverging in every
coordinate.
\end{itemize}
and if {\rm (i)--(iv)} do not hold, then $(U_n(x_E(1),y_E(1),z_E(1)))_{n \in \N}$ is bounded by
$C(E)$, where the constant $C(E)$ depends continuously on $E$.
\end{prop}

\begin{proof}
It follows from \eqref{qvrep} and \eqref{undef} that $U_n$ can be written as a product of blocks of
type $vq^{\pm 1}$. The strategy proposed by Roberts is to study the evolution of the initial vector
$(x_E(1),y_E(1),z_E(1))$ under these elementary blocks, that is, to enlarge the orbit in question.
This enlarged orbit will be denoted by $(\xi_m)_{m \in \N}$, where $\xi_m$ is the vector obtained after
applying the first $m$ blocks of type $v q^{\pm 1}$ to $(x_E(1),y_E(1),z_E(1))$. It was shown in
\cite{r} that if $\xi_M \in \mathcal{E}$ for some $M \in \N$, then $\xi_m \in \mathcal{E}$ for every $m
\ge M$ and $\xi_m$ diverges super-exponentially in every coordinate. This readily gives $(ii)
\Rightarrow (iii) \Rightarrow (iv)$. Of course, $(iv) \Rightarrow (i)$ is obvious. 

To show that $(i) \Rightarrow (ii)$, we need some preparation.
Let $b_i \in \{vq, vq^{-1}\}$, $i \in \N$, be chosen such that
$\ldots G_3 G_2 G_1 = \ldots b_5 b_4 b_3 b_2 b_1$ and define $\xi_m = b_m \ldots b_1 (x,y,z)$. Since
$q,q^{-1},v$ preserve the invariant $I=I(x_E(1),y_E(1),z_E(1))$, we have $I(\xi_m) = I$ for every $m
\in \N$. It is easily seen  (cf.\ \cite{r}) that, if

\begin{equation}\label{outside}
\| \xi_m \|_2^2 > 2 + (1 + \sqrt{I})^2,
\end{equation}
$\xi_m$ has at least two coordinates whose modulus is greater that one, where $\| \cdot \|_2$ denotes
the norm $\|(x,y,z)\|_2^2 = |x|^2 + |y|^2 + |z|^2$. Thus every $\xi_m =(\xi^1_m,\xi^2_m,\xi^3_m)$
obeying \eqref{outside} has to be of one of the following forms,

\begin{equation}\label{altern}
\begin{array}{lrrr}
{\rm (I)} & |\xi^1_m| > 1, & |\xi^2_m| > 1, & |\xi^3_m| > 1,\\
{\rm (II)} & |\xi^1_m| \le 1, & |\xi^2_m| > 1, & |\xi^3_m| > 1,\\
{\rm (III)} & |\xi^1_m| > 1, & |\xi^2_m| \le 1, & |\xi^3_m| > 1,\\
{\rm (IV)} & |\xi^1_m| > 1, & |\xi^2_m| > 1, & |\xi^3_m| \le 1.
\end{array}
\end{equation}
If a $\xi_m$ satisfies \eqref{outside}, we say it is of type I, II, III, or IV if it satisfies the
respective condition in \eqref{altern}.

\medskip

Let us now describe a scenario sufficient for entry in $\mathcal{E}$ (cf.  \cite{d3,r} for similar
reasoning). Suppose $\xi_{m_0}$ obeys \eqref{outside} and

\begin{equation}\label{xiinc}
\| \xi_{m_0 + 1} \|  = \| v q^\varepsilon \xi_{m_0} \| > \| \xi_{m_0} \|,
\end{equation}
where $\varepsilon \in \{ 1, -1 \}$. We  consider the case $\varepsilon = 1$, the other case is
similar.   If $\xi_{m_0}$ is of type I, II, or IV, then $\xi_{m_0 + 1}$ is in $\mathcal{E}$. This
follows by simple direct calculations considering the cases $\xi_{m_0}\in \mathcal{E}$ and  
$\xi_{m_0}\notin \mathcal{E}$ (cf.\ Proposition 3.4 in \cite{r} as well).  If $\xi_{m_0}$ is of type
III, then we proceed as follows (cf.\ \cite{d3}). The iterate faces $vq^{-1} (vq)^j$ for some
$j \ge 1$. If $\xi_{m_0 + j}$ obeys \eqref{outside}, then it must be one of the types I--IV. It is easy
to see that $qv (\xi)$ fixes the second component of $\xi$. Thus, $\xi_{m_0 + j}$ must be of type III
as well. Using this, it is straightforward to  see that $\xi_{m_0 + j + 1}$ belongs to $\mathcal{E}$.
Thus, we infer that indeed the scenario

\begin{itemize}
\item[(S)] $ \| \xi_{m_0 + 1} \|  = \| v q^\varepsilon \xi_{m_0} \| > \| \xi_{m_0} \|$  and
$\|\xi_{m_0}\|, \|\xi_{m_0 + j}\|\geq (2 + (1 + \sqrt{I})^2)^{1/2}$
\end{itemize}
forces entry in $\mathcal{E}$.

\medskip

Our next step is to investigate  how    growing of $U_n$ (or $V_n$) can force (S).  We only consider
$U$.  Consider the situation

\begin{equation}\label{ungrows}
\|U_{n+1} (x_E(1), y_E(1), z_E(1))\| > \|U_n (x_E(1), y_E(1), z_E(1))\|.
\end{equation}
Thus we have at least one situation of the form \eqref{xiinc} between the two iterates and among
these, we consider the one with largest $m_0$ for definiteness. To make sure that $\xi_{m_0}$ obeys
\eqref{outside} it suffices to assume that $U_{n+1} (x_E(1), y_E(1), z_E(1)) = (x_E(2n+3), y_E(2n+3),
z_E(2n+3))$ has norm sufficiently large. More precisely, let $R$ be given with
$$
\|v q (x)\|, \|v q^{-1} (x)\| > R
$$
whenever $\|x\| \leq (2 + (1 + \sqrt{I})^2)^{1/2}$. Then \eqref{ungrows} together with
$\|U_{n+1}\|\geq R$ forces that $\xi_{m_0}$ obeys \eqref{outside}. To make sure that $\xi_{m_0 + j}$
obeys \eqref{outside}, note that 
$$
\xi_{m_0 + j}= p^{-1} F_{2n+1} (U_n (x_E(1), y_E(1), z_E(1))) = p^{-1} V_n (x_E(2), y_E(2), z_E(2))
$$
if $\xi_{m_0 + j}=(v q)^j \xi_{m_0}$ and  
$$
\xi_{m_0 + j}= U_{n+1} ( (x_E(1), y_E(1), z_E(1))
$$
if $\xi_{m_0 + j}=(v q^{-1})^j \xi_{m_0}$ respectively. Thus, as $p$ is an isometry,  $\xi_{m_0 + j}$
obeys \eqref{outside} if $ V_n (x_E(2), y_E(2), z_E(2))$ and $ U_{n+1} ( (x_E(1), y_E(1), z_E(1)) $ 
have  norm larger than $(2 + (1 + \sqrt{I})^2)^{1/2}$. 

Let us summarize these considerations as follows:  \eqref{ungrows} together with sufficient largeness
of  $\| U_{n+1} (x_E(1), y_E(1), z_E(1))\|$ and $ \| V_n (x_E(2), y_E(2), z_E(2))\|$ force (S) and thus
entry in the ecape set. A  completely analoguous arguments  applies  after interchanging $V$ and $U$.

\medskip

To conclude the proof, we show now that if (ii) fails, then (i) fails with explicit upper bound on the
orbit. This will prove both $(i) \Longrightarrow (ii)$ and the last statement of the Proposition.  

Thus, assume that (ii) fails for some energy $E$ and write $U_m$ for $U_m (x_E(1), y_E(1), z_E(1)) =
(x_E(2m+1), y_E(2m+1), z_E(2m+1))$ and $V_m$ for $V_m (x_E(2), y_E(2), z_E(2)) = (x_E(2m+2), y_E(2m+2),
z_E(2m+2))$.  By definition of $U_n$, $V_n$ and the invariance of $I$, we have

\begin{equation}\label{nachbargross}
\mbox{ $\|U_{n+1}\|$ large  implies   $\|V_n\|$ or $\|V_{n+1}\|$ is larger than $ R  (2 + (1 +
\sqrt{I})^2)^{1/2}$.}
\end{equation}
(A similar argument is used in the proof of Corollary \ref{globalbound}.) Now, we have  for every $n$
where \eqref{ungrows} occurs with $\|U_{n+1}\|$ large enough  for \eqref{nachbargross} to hold  the
following situation: If $\|V_n\|$ is large, then we have a scenario sufficient for entry in
$\mathcal{E}$ as explained above, which is a contradiction to failure of (ii). If
$\|V_{n+1}\|$ is large (but $\|V_n\|$ is not), then we consider the extended orbit between $V_n$ and
$V_{n+1}$ and we are again in a scenario sufficient for entry in $\mathcal{E}$ because $\xi_{m_0}$ can
be chosen such that the role of the intermediate vector $\xi_{m_0 + j}$ is played by $p^{-1} U_{n+1}$
whose norm is large by assumption, again a contradiction. This shows that if (ii) fails, we get an
explicit upper bound for $\|U_{n+1}\|$ in all situations where we have \eqref{ungrows}. In particular,
(i) fails and we have an upper bound on the $U_n$--orbit in this case. Note that the upper bound we get
is an explicit function of $I=I(x_E(1),y_E(1),z_E(1))$ and $\|(x_E(1),y_E(1),z_E(1))\|$ and hence is
continuous in $E$.
\end{proof}

Define the stable set 

\begin{equation}
\mathcal{B} = \left\{ E \in \R : \sup_{n \in \N} \|(U_n(x_E(1), y_E(1) ,z_E(1))\| \le C(E) \right\}.
\end{equation}

\begin{prop}\label{sinb}
$\Sigma \subseteq \mathcal{B}$.
\end{prop}

\begin{proof}
Recall from \eqref{spectrum} that $\Sigma$ is equal to $\sigma(H_\omega)$ for every $\omega \in
\Omega$. It follows from a standard strong approximation procedure that 

\begin{equation}\label{strong1}
\bigcup_{k \in \N} {\rm Int} \left( \bigcap_{m \ge k} \{ E \in \R : |y_E(2m+1)| > 1 \} \right)
\subseteq \Sigma^c.
\end{equation}
In fact, the set $ \{ E \in \R : |y_E(n)| > 1 \} $ is just $\R \setminus \sigma(H_{n})$ where
$H_{n}$ is a $|s_{n}'|$-periodic operator whose potential values over one period are given by
$f(s_{n}')$. Clearly, any operator $H_\omega$ is a strong limit of such operators, so
we can apply Theorem VIII. 24 of ~\cite{rs1} to get~\eqref{strong1}.

Using Proposition~\ref{orbitchar} and \eqref{strong1}, we get the following chain
of inclusions:

\begin{eqnarray*}
\mathcal{B}^c & = & \bigcup_{n \in \N} \{ E : U_n(x_E(1), y_E(1) ,z_E(1)) \in \mathcal{E} \}\\
& = & \bigcup_{n \in \N} {\rm Int} \left( \{ E : U_n(x_E(1), y_E(1) ,z_E(1)) \in \mathcal{E} \}
\right)\\ & = & \bigcup_{n \in \N} {\rm Int} \left( \bigcap_{m \ge n} \{ E : U_m(x_E(1), y_E(1)
,z_E(1)) \in \mathcal{E} \} \right)\\ & \subseteq & \bigcup_{n \in \N} {\rm Int} \left( \bigcap_{m \ge
n} \{ E : |y_E(2m+1)| > 1 \} \right)\\
& \subseteq & \Sigma^c
\end{eqnarray*}
\end{proof}

We now relate the stable set $\mathcal{B}$ to the set $A$ of energies $E$ for which the Lyapunov
exponent $\gamma(E)$ vanishes. Let us briefly recall the definition of $\gamma(E)$. If $\Omega =
\Omega_u$ is a quasi-Sturmian hull with potentials $V_\omega$, $\omega \in \Omega$, then we define
transfer matrices
$$
M_{\omega,E} (n) = \left( \begin{array}{cc} E - V_\omega(n) & -1 \\ 1 & 0 \end{array} \right) \times
\cdots \times \left( \begin{array}{cc} E - V_\omega(1) & -1 \\ 1 & 0 \end{array} \right).
$$
If $\mu$ denotes the unique ergodic measure on $\Omega$, it follows from the subadditive ergodic
theorem that for every $E \in \C$, there exists a nonnegative number $\gamma(E)$ such that for almost
every $\omega \in \Omega$ with respect to $\mu$,

\begin{equation}\label{gammadef}
\gamma(E) = \lim_{n \rightarrow \infty} \frac{1}{n} \ln \| M_{\omega,E}(n) \|.
\end{equation}
Let
$$
A = \{ E \in \R : \gamma(E) = 0 \}.
$$

The following proposition shows that a bounded trace map orbit implies vanishing Lyapunov exponent.
The proof is modelled after \cite{d3} and is considerably shorter than the proof given in \cite{bist}
for the Sturmian case (cf. \cite{dl4} for related material).

\begin{prop}\label{bina}
$\mathcal{B} \subseteq A$.
\end{prop}

\begin{proof}
Assume there exists $E \in \mathcal{B}$ such that $\gamma(E) > 0$. Pick some $\omega \in \Omega$ for
which the limit in \eqref{gammadef} exists. By Osceledec's theorem \cite{cfks} there exists a solution
$\phi_+$ of $H_\omega \phi = E \phi$ such that $\|\Phi_+(m)\|$ decays exponentially at $+\infty$ at the
rate $\gamma(E)$, where $\Phi_+(m)=(\phi_+(m+1),\phi_+(m))^T$. Now, it follows from \eqref{casseq} and
the fact that the words $s_n s_n$ occur infinitely often in $u_{{\rm St}}$ \cite{b,dl1} that the words
$s_n' s_n'$, $n \in \N$, occur infinitely often in $u$ and hence in $\omega$. Since $E \in \mathcal{B}$
there is a constant $C \ge 1$ such that, for every $n \in \N$, we have

\begin{equation}\label{tb}
|{\rm tr}(M_E(n))| \le C.
\end{equation}
Pick $m_0$ such that, for every $m \ge m_0$ and every $k \in \N$, the solution $\phi_+$ obeys

\begin{equation}\label{decay}
\|\Phi_+(m+k)\| \le \exp (-\tfrac{1}{2}\gamma(E)k) \|\Phi_+(m)\|.
\end{equation}
Choose $n$ such that $\exp (-\frac{1}{2} \gamma(E) |s_n'|) < \frac{1}{2C}$. Look for an occurrence of
$s_n' s_n'$ in $\omega$, that is,  $s_n' s_n' = \omega(l+1)\ldots \omega(l+2|s_n'|)$, such that $l \ge
m_0$. It follows from the Cayley-Hamilton theorem that

\begin{equation}\label{quadr}
\Phi_+(l+2|s_n'|) - {\rm tr} (M_E(n)) \Phi_+(l+|s_n'|) + \Phi_+(l) = 0,
\end{equation}
which in turn implies by \eqref{tb}

\begin{equation}\label{nondec}
\max(\|\Phi_+(l+|s_n'|)\|, \|\Phi_+(l+2|s_n'|)\|)  \ge \frac{1}{2C}  \|\Phi_+(l)\|,
\end{equation}
contradicting (\ref{decay}).  
\end{proof}

\begin{prop}\label{ains}
$A \subseteq \Sigma$.
\end{prop}

\begin{proof}
This is well known \cite{cl}.
\end{proof}

We collect the results of Propositions~\ref{sinb} through \ref{ains} in the following corollary which
provides an extension of the main theorem of \cite{bist} to the quasi-Sturmian case. 

\begin{coro}\label{chaineq}
$\Sigma = \mathcal{B} = A$.
\end{coro}

Since the spectrum is compact and the bounds on the trace map orbit for energies $E$ in the spectrum
depend continuously on $E$, we can find a global bound for these orbits. This observation will be
important in Section~\ref{alpha}, so we state the following corollary.

\begin{coro}\label{globalbound}
There is a constant $C = C(u,f) < \infty$ such that for every $E \in \Sigma$ and every $n \in \N$, we
have
$$
\max \{ \| \tr(M_E(n)) \| , \| \tr(M_E(n) M_E(n-1)) \| \} \le C.
$$
\end{coro}

\begin{proof}
As explained above, Corollary~\ref{chaineq} and compactness of $\Sigma$ provide uniform bounds on
$$
\| U_n (x_E(1), y_E(1), z_E(1)) \| = \| (x_E(2n+1), y_E(2n+1), z_E(2n+1))\|
$$
for $E \in \Sigma$, $n \in \N$, that is, uniform bounds on
$$
\|(x_E(n), y_E(n), z_E(n))\|
$$
for every odd $n$ and every $E \in \Sigma$. By $y_E(n) = x_E(n+1)$ this gives uniform bounds on
$\|x_E(n)\|$ and $\|y_E(n)\|$ for \textit{every} $n$ and we can then use the invariant $I$ to derive a
uniform bound for $\|z_E(n)\|$, $E \in \Sigma$ and every $n \in \N$.
\end{proof}

\section{Zero-Measure Spectrum and Absence of Absolutely Continuous Spectrum}\label{ac}

Zero-measure spectrum and absence of absolutely continuous spectrum follows immediately from
Corollary~\ref{chaineq} and Kotani~\cite{k}; just as in the Sturmian case \cite{bist} (see also
\cite{s2}). Namely, Kotani has shown the following: Given an ergodic family of discrete one-dimensional
Schr\"odinger operators with aperiodic potentials taking finitely many values, we always have

\begin{equation}\label{kotani}
|A| = 0,
\end{equation}
where $A$ is the set of energies for which the Lyapunov exponent vanishes and $| \cdot |$ denotes
Lebesgue measure. These assumptions hold in our present context, so combining this equality with
Corollary~\ref{chaineq}, we get that the spectrum has zero Lebesgue measure. Moreover, since a set of
zero Lebesgue measure cannot support absolutely continuous spectrum, we also get absence of absolutely
continuous spectrum for any quasi-Sturmian potential. More generally, uniform absence of absolutely
continuous spectrum can also be proved in the following way: It follows from \eqref{kotani} that almost
every operator in the ergodic family has empty absolutely continuous spectrum and since the hull is
minimal in the quasi-Sturmian case this extends to \textit{all} elements of the family by a result of
Last and Simon \cite{ls}. We can therefore state the following:

\begin{prop}\label{acabsence}
Let $u$ be quasi-Sturmian and let $f$ be one-to-one. Then for every $\omega \in \Omega$, the operator
$H_\omega$ has empty absolutely continuous spectrum and its spectrum as a set has zero Lebesgue measure.
\end{prop}

\section{Absence of Point Spectrum}\label{pp}

In this section we prove absence of eigenvalues for all operators with quasi-Sturmian potentials.
Sections~\ref{ac} and \ref{pp} thus imply Theorem~\ref{main}. Proposition~\ref{cass} provides us with
an excellent tool to carry over the approach of \cite{dl1} (cf.\ \cite{len2} as well) and extend the
result of \cite{dkl,dl1} to the quasi-Sturmian case.

We will prove the following result which, together with Proposition~\ref{acabsence}, implies
Theorem~\ref{main}.

\begin{prop}\label{ppabsence}
Let $u$ be quasi-Sturmian and let $f$ be one-to-one. Then for every $\omega \in \Omega$, the operator
$H_\omega$ has empty point spectrum.
\end{prop}

To prove Proposition~\ref{ppabsence} we pursue a similar strategy as in \cite{dkl,dl1}. The criterion
for excluding eigenvalues is Gordon's two-block method~\cite{d2,g} which can be phrased as
follows~\cite{dl1}.

\begin{lemma}\label{gordon}
Let $V$ be a two-sided sequence such that for some $m \in \Z$, $V$ has infinitely many squares $w_n
w_n$ starting at $m$. If, for some energy $E \in \R$, the traces of the transfer matrices over the
blocks $w_n$ are bounded, $E$ is not an eigenvalue of the discrete Schr\"odinger operator
$$
(H \phi) (n) = \phi(n+1) + \phi(n-1) + V(n) \phi(n).
$$
\end{lemma}

We will show that sufficiently many squares can be found for every $\omega$ in a quasi-Sturmian family:

\begin{lemma}\label{gordonworks}
Let $u$ be quasi-Sturmian. Then for every $\omega \in \Omega$, there is a site $m$ such that the
sequence $\omega$ has infinitely many squares conjugate to some $s_n'$ or some $s_n' s_{n-1}'$ starting
at $m$.
\end{lemma}

\begin{proof}
Let $u \in \mathcal{A}^\N$ be a quasi-Sturmian sequence and let $\omega \in \Omega$. Since $\omega$ is
necessarily recurrent, its restriction $\omega|_\N$ to $\N$ has the form \eqref{casseq} for some
Sturmian $u_{{\rm St}} \in \{a,b\}^\N$, some aperiodic substitution $S:\{a,b\}
\rightarrow \mathcal{A}$, and some finite word $w \in \mathcal{A}^*$; see the appendix and
Proposition~\ref{rotnum} in particular for details. Let $m = |w|$. It was shown in \cite{dkl} that
$u_{{\rm St}}$ starts with infinitely many squares conjugate to $s_n$ or $s_n s_{n-1}$. Thus the lemma
follows immediately from \eqref{casseq} and \eqref{sn'}.
\end{proof}

\begin{proof}[Proof of Proposition~\ref{ppabsence}]
Using Corollary~\ref{globalbound} and Lemma~\ref{gordonworks}, we see that Lemma~\ref{gordon} applies
to $V_\omega$ for every $\omega \in
\Omega$ and every $E \in \Sigma = \sigma(H_\omega)$. Thus for every $\omega \in \Omega$, the operator
$H_\omega$ has empty point spectrum.
\end{proof}

We end this section with a brief discussion of this result. For operators \eqref{oper} with potential
$V$ taking finitely many values, the standard approaches to an ``absence of eigenvalues''--type result
employ either powers or palindromes. We have shown in this section that a Gordon criterion, which is
based on powers (the occurrence of infinitely many squares), is applicable to quasi-Sturmian models and
it gives uniform (i.e., for every $\omega \in \Omega$) absence of eigenvalues. In the Sturmian case,
the palindrome method is also applicable, as shown by Hof et al.\ in \cite{hks}, but is gives a weaker
result: Absence of eigenvalues can be established for a dense $G_\delta$~set of $\omega$'s in $\Omega$.
For some quasi-Sturmian models, however, the palindrome method does not apply at all! Namely, if we
take an arbitrary Sturmian sequence $u_{{\rm St}} \in \{a,b\}^\N$ and apply the aperiodic substitution
$S(a) = 011001$, $S(b) = 001011$, we obtain a quasi-Sturmian sequence $u$ which is not palindromic (see
\cite{abcd}) in the sense of \cite{hks}, so the method of Hof et al.\ does not apply!

\section{{\larger[2]$\alpha$}-Continuity}\label{alpha}

In this section we prove Theorem~\ref{acont}. Corollary~\ref{globalbound} provides us with the crucial
tool, namely, uniformly bounded trace map orbits for energies from the spectrum. It has been
demonstrated in~\cite{dkl} that $\alpha$-continuity of a whole-line operator can be shown by proving
power-law upper and lower bounds for local $\ell^2$-norms of generalized eigenfunction on a half line. 

The lower bound can be established by a technique similar to the one leading to absence of eigenvalues
in Section~\ref{pp}. Thus, given boundedness of trace map orbits, it  can be proved in the same way as
in the Sturmian case. The proof of the upper bound extends the works \cite{dl2,irt,len}. This requires
some extra care as the underlying shifts are not symmetric under reflection. To overcome this
difficulty we use the results discussed at the end of Section \ref{models}. For further information and
related material we refer the reader to 
\cite{dkl,dl2,irt,len}.

Let us first recall a criterion for $\alpha$-continuity of a whole-line operator from~\cite{dkl}. It
complements the half-line results of Jitomirskaya and Last which were established in~\cite{jl1}. Given
an operator $H$ as in \eqref{oper}, we consider the solutions of $H \phi = E\phi$, that is, solutions
to the difference equation

\begin{equation}\label{eve}
\phi(n+1) + \phi(n-1) + V(n) \phi(n) = E \phi (n).
\end{equation}
We define their local $\ell^2$-norm by 

\begin{equation}\label{lnorm}  
\|\phi\|_L^2 = \sum_{n=0}^{\lfloor L \rfloor} \big|\phi(n)\big|^2 \; + \;(L-\lfloor L
\rfloor)\big|\phi(\lfloor L \rfloor +1)\big|^2.
\end{equation}
We call a solution $\phi$ of \eqref{eve} normalized if $|\phi(0)|^2 + |\phi(1)|^2 = 1$. The following
result was proved in~\cite{dkl}.

\begin{prop}\label{gcd}
Let $\Sigma$ be a bounded set. Suppose there are constants $\gamma_1, \gamma_2$ such that for each
$E\in\Sigma$, every normalized solution of
\eqref{eve} obeys the estimate

\begin{equation}\label{powerbounds}    
C_1(E) L^{\gamma_1} \leq \|u\|_L \leq C_2(E) L^{\gamma_2}  
\end{equation}
for $L>0$ sufficiently large and suitable constants $C_1(E), C_2(E)$. Let $\alpha ={2
\gamma_1}/({\gamma_1 + \gamma_2})$. Then $H$ has purely $\alpha$-continuous spectrum on $\Sigma$, that
is, for any $\phi\in\ell^2$, the spectral measure for the pair $(H,\phi)$ is purely $\alpha$-continuous
on $\Sigma$. Moreover, if the constants $C_1(E), C_2(E)$ can be chosen independently of $E \in \Sigma$,
then for any $\phi\in\ell^2$ of compact support, the spectral measure for the pair $(H,\phi)$ is
uniformly $\alpha$-H\"older continuous on $\Sigma$.
\end{prop}

We want to show that the bounds \eqref{powerbounds} can be established for every potential $V_\omega$
associated with some quasi-Sturmian sequence $u$ provided that the rotation number has bounded density.
In fact, the constants $C_i, \gamma_i$ can be chosen uniformly in $\omega \in \Omega$ and $E \in
\Sigma$, so we get that for any $\omega \in \Omega$ and any $\phi\in\ell^2$ of compact support, the
spectral measure for the pair $(H_\omega,\phi)$ is uniformly $\alpha$-H\"older continuous.

We proceed similarly to the proof of Lemma~\ref{gordonworks}. Namely, we consider an arbitrary element
$\omega$ of the hull $\Omega$ and the associated operator $H_\omega$. Since we are interested in the
behavior of the solutions of

\begin{equation}\label{omeve}
\phi(n+1) + \phi(n-1) + V_\omega(n) \phi(n) = E \phi (n)
\end{equation}
on the right half-line, we consider the restriction of $\omega$ to the right half-line and its
representation as in~\eqref{casseq}. Using this representation and properties of the associated
Sturmian sequence (whose rotation number is $\theta_c$ and has bounded density; compare
Proposition~\ref{rotnum}), along with the trace map bounds from Corollary~\ref{globalbound}, we can
obtain the following two propositions.

\begin{prop}\label{lowerpower}  
Let $\theta_c$ be such that for some $B < \infty$, the denominators $q_n$ of the associated rational
approximants obey $q_n \le B^n$ for every
$n \in \N$. Then for every injective $f$, there exist $0 < \gamma_1, C_1 < \infty$ such that for every
$E \in \Sigma$ and every $\omega \in
\Omega$, every normalized solution $\phi$ of \eqref{omeve} obeys  

\begin{equation}\label{lpb}  
\| \phi \|_L \ge C_1 L^{\gamma_1}  
\end{equation}  
for $L$ sufficiently large.  
\end{prop}  

\noindent\textit{Remark.} The set of $\theta_c$'s obeying the assumption of Proposition
\ref{lowerpower} has full Lebesgue measure \cite{khin} and clearly contains the set of bounded density
numbers.

\begin{proof}
Using Corollary~\ref{globalbound} and Proposition~\ref{rotnum} this can be shown in the same way as in
the Sturmian case. Details can be found in \cite{dkl} (cf.\ \cite{d1} as well). 
\end{proof}

\begin{prop}\label{upperpower}  
Let $\theta_c$ be a bounded density number. Then for every injective $f$, there exist $0 < \gamma_2, C_2 < \infty$ such that for every $E \in
\Sigma$ and every $\omega \in \Omega$, every normalized solution $\phi$ of \eqref{omeve} obeys  

\begin{equation}\label{upb}  
\| \phi \|_L \le C_2 L^{\gamma_2}  
\end{equation}  
for all $L$.  
\end{prop}

\begin{proof}
The proof is similar to the proof of the corresponding result in the Sturmian case (cf.\
\cite{dkl,dl2,len}). However, it requires some additional effort as the quasi-Sturmian systems are not
reflection invariant. To treat this case as well we will need the results (and the notation) of the
discussion at the end of Section \ref{models}.

Recall that $\Omega=\Omega(\theta,S)$. Mimicking the argument of \cite{irt}, which only relies on
trace map bounds and recursions, one can easily infer that there exists $C_1$ and $\gamma_1$ with
$$
\| M(S(y),E)\|\leq C_1 |y|^{\gamma_1}
$$
for every prefix $y$ of $c_\theta$ and every $E\in \Sigma(\Omega(\theta,S))$. Now, every prefix of
$S(c_{\theta})$ can be written as $x= S(y) w$ with a prefix $y$ of $c_\theta$ and a word $w$ of length
$|w|\leq \max\{ |S(0)|, |S(1)|\}$. Combining these estimates, we see that there exist $C$ and $\gamma$
with

\begin{equation}\label{bound}
\|M(x,E)\|\leq C |x|^\gamma\:\;\mbox{  for every prefix $x$ of $S(c_{\theta})$ and every $E \in
\Sigma(\Omega(\theta,S))$ }.
\end{equation}
By, Corollary \ref{invarianz}, we have $\Sigma(\Omega(\theta,S))= \Sigma(\Omega(\theta,S^R))$. Thus, 
replacing $S$ by $S^R$ but keeping the \textit{same} set of energies,  we  can again apply the above
reasoning and  infer
\begin{equation}\label{boundzwei}
\|M(x,E)\|\leq C |x|^\gamma\:\;\mbox{  for every prefix $x$ of $S^R (c_{\theta})$ and every $E \in
\Sigma(\Omega(\theta,S))$}.
\end{equation}

Now, the proof can be finished as follows. By results of \cite{dl2,len}, every factor $z$ of
$c_\theta$ can be written as $z=x y$, where $x$ is a suffix of a suitable $s_n$ and $y$ is a prefix of
$s_{n+1}$. This implies that every factor $z$  of a sequence in $\Omega(\theta,S)$ can be written as $z
= x y$ with $x$ a suffix of $S(s_n)$ and $y$ a prefix of $S(s_{n+1})$ for a suitable $n$. Assume
w.l.o.g.\ $n\geq 3$. (The case $n=1,2$ can be treated directly.)  By equation
\eqref{palin}, $s_n= \pi_n a b$ with $a,b\in \{0,1\}$ and $a\neq b$. 

We will assume that $|x|> |S(a) S(b)|$. The other case is similar (and, in fact, even simpler). Then
$x$ can be written as $x= \widetilde{x} S(a) S(b)$, where $\widetilde{x}$ is a suffix of $S(\pi_n)$.
Now, we can estimate
$$
\| M( z ,E)\|\leq \|M(S(a),E)\| \cdot \|M(S(b),E)\| \cdot \|M(\widetilde{x},E)\|\cdot \|M(y,E)\|.
$$
We will provide suitable bounds for all these factors. The factor $ \|M(S(a),E)\| \|M(S(b),E)\|$ is
just a constant. The factor $ \|M(y,E)\|$ can be estimated by equation \eqref{bound} as $y$ is a prefix
of$S(s_{n+1})$. We find $\|M(y,E)\| \leq C |y|^\gamma\leq C |z|^\gamma$. 

It remains to estimate the factor $ \|M(\widetilde{x},E)\|$. As $\widetilde{x}$ is a suffix of
$S(\pi_n)$, the word  $\widetilde{x}^R$ is a prefix  of $S(\pi_n)^R$. Since $\pi_n$ is a palindrome, we
have $S(\pi_n)^R=S^R (\pi_n)$. This means that $\widetilde{x}^R$ is a prefix of
$S^R(c_\theta)$. By equation \eqref{boundzwei}, we can thus estimate $\|M(\widetilde{x}^R,E)\|$. By
general principles, however, we have $
\|M(\widetilde{x},E)\|= \|M(\widetilde{x}^R,E)\|$ (cf.\ \cite{dl2,len}) and we see that
$\|M(\widetilde{x},E)\|$ can be estimated by $\|M(\widetilde{x},E)\|\leq C |\widetilde{x}|^\gamma\leq C
|z|^\gamma$. 

Combining these estimates we arrive at the desired statement.  
\end{proof}

\begin{proof}[Proof of Theorem~\ref{acont}]
The claim follows from Propositions~\ref{lowerpower} and~\ref{upperpower} along with
Poposition~\ref{gcd}.
\end{proof}

\medskip

\noindent\textit{Acknowledgements.} A substantial part of this work was done while one of the authors (D.~L.) was visiting The Hebrew University, Jerusalem.
He would like to thank Y. Last for hospitality and many useful discussions.

\section{Appendix: Rotation numbers of quasi-Sturmian sequences}

In this section we discuss Cassaigne's proof of Proposition~\ref{cass} and its consequences for the rotation numbers associated with the
elements in a quasi-Sturmian hull. We show in particular that given a quasi-Sturmian sequence $u$, every rotation number for $u$ is a rotation
number for every element of the associated subshift $\Omega$ and vice versa.

Let $u$ be quasi-Sturmian, that is, $u$ is recurrent and satisfies $p_u(n) = n+k$ for $n \ge n_0$ and let $\Omega$ be the induced subshift
containing two-sided sequences whose factors are also factors of $u$. It is well known that $u$ must be uniformly recurrent so that each
factor of $u$ is also a factor of every $\omega \in \Omega$ and hence all elements of $\Omega$ are factor-equivalent. We will show that the
set of allowed rotation numbers is an invariant of the subshift. 

To do so, let us briefly recall how the Sturmian sequence $u_{{\rm St}}$ in Proposition~\ref{cass} is found. The Rauzy graphs $G(n)$, $n \in \N$
associated with $u$ are defined as follows: For every $n \in \N$, the graph $G(n)$ contains $p_u(n)$ vertices corresponding to the factors of
$u$ having length $n$ and it has $p_u(n+1)$ edges corresponding to the factors of $u$ having length $n+1$. There is an edge from $w_1$ to $w_2$
if and only if $w_1 = ax$, $w_2 = xb$, $|a| = |b| = 1$ ($|x| = n-1$), and $axb$ is a factor of $u$. Since $u$ is quasi-Sturmian, for large
enough $n$, the topology of $G(n)$ is the same as the topology of a Rauzy graph of a Sturmian sequence. Namely, $p_u(n+1) - p_u(n) = 1$ and
there is a unique right-special factor (a vertex with out-degree $2$) and a unique left-special factor (a vertex with in-degree $2$).
Similarly, one may define Rauzy graphs for every $\omega \in \Omega$. Observe that the family of Rauzy graphs is the same for $u$ and every
$\omega \in
\Omega$. For certain $n$, it can be shown that the right-special factor and the left-special factor coincide (i.e., there exists a bispecial
factor). Fix such an $n$ and consider the graph $G(n)$. It has a vertex with in-degree $2$ and out-degree $2$ and exactly two paths starting
and ending at this vertex. Each vertex in these two paths (other than the vertex $B$ corresponding to the bispecial factor) has in-degree $1$
and out-degree $1$ and one finds the Sturmian sequence $u_{{\rm St}}$ by running through $u$ (or one of the $\omega \in \Omega$) and recording
the sequence of choices as to which of the two paths is taken after every passage through the vertex $B$. The prefix $w$ in
Proposition~\ref{cass} is just the word we have to read to get from the starting site to the first occurrence of the bispecial factor.

From the above remarks, we can deduce the following. Since $u$ and all $\omega \in \Omega$ are factor-equivalent, the Sturmian sequences
$u_{{\rm St}}$ that can be obtained must be factor-equivalent as well. This shows that the rotation numbers we can get this way must be valid
for all of these sequences. Moreover, the set of rotation numbers is naturally labelled by the set of integers $n$ for which there exists a
bispecial factor. For definiteness, we may choose the rotation number corresponding to the shortest bispecial factor as the \textit{canonical
rotation number} associated with $\Omega$.

We summarize our observations in the following proposition.

\begin{prop}\label{rotnum}
Let $u$ be a quasi-Sturmian sequence. Then there is a canonical rotation number $\theta_c$ which is a rotation number for all the elements of
the subshift $\Omega$. By restricting the elements $\omega \in \Omega$ to the right half-line, we obtain one-sided infinite words that all
have a representation of the form \eqref{casseq}. In this representation, one can accomplish the following: The substitution $S$ is
independent of $\omega$, while the Sturmian sequence $u_{{\rm St}} = u_{{\rm St}}(\omega)$ does depend on $\omega$, however, its rotation number
is $\theta_c$ and hence independent of $\omega$. The prefixes $w = w(\omega)$ depend on $\omega$ and they are suffixes of some $s_n'$.
\end{prop}

\end{document}